\documentclass[floatfix,showpacs,showkeys,amsmath,preprintnumbers,nofootinbib,twocolumn] {revtex4}
\usepackage{graphicx}
\usepackage{colordvi}
\usepackage{amsmath}
\usepackage{amssymb}
\usepackage{latexsym}
\usepackage{hyperref}
\begin{document}
\title{Effect of Interacting Dark Energy on Expansion of the  Universe}

\author{Debasis Sahu$^*$ and Bibekananda Nayak$^\dagger$ }

\address{P.G. Department of Applied Physics and Ballistics, Fakir Mohan University,
Balasore, Odisha 756019, India\\
E-mail: $^*$debasissahu777@gmail.com, $^\dagger$bibekanandafm@gmail.com}
\begin{abstract}
In our present work, we study the evolution of the universe by
assuming an interacting dark energy model, where dark energy interacts with matter. Basing on this model, first we calculated the dark
energy density parameter and using this we have picturised the expansion of the universe. From our analysis, we found that
presently observed accelerated expansion of the universe can be explained by interacting model, if the dark energy is quintessence type. 
Though equation of sate parameter of dark energy $\gamma_{\phi}$ for quintessence varies between $0$ and $-1$, our results predict that accelerated expansion is only possible for $\gamma_{\phi}$ less  (more negative) than $-0.166$. It is also found that in early time the universe was undergoing an decelerated phase of expansion and transition from deceleration to acceleration would occur in recent past. Further, our model predicts that in near future again expansion of the universe will undergo a  second transition from accelerated phase to a decelerated one and finally deceleration parameter will take a constant positive value $1/2$ as early universe indicating a constant rate of deceleration in far future like distant past.
\end{abstract}



\maketitle
\section{Introduction}	

Along with Friedman-Robertson-Walker metric which describes a homogeneous and isotropic universe, Einstein's field equation builds the backbone of Standard Model of Cosmology. This model is not only supported by Hubble's observation of expanding universe but also by many other astrophysical and extra-galactic phenomena including recently observed gravitational wave. But it predicts that universe would undergo a decelerated expansion through out its evolution where as some of the current observations \cite{1,2} show that the present universe is expanding at an accelerated rate. For the removal of this controversy, in theoretical cosmology, generally two ways have been adopted. One is by modifying the theory of gravity \cite{3,4} and the other is by introducing an unknown form of energy having negative pressure coined as dark energy \cite{5,6,7}. The dark energy is homogeneous, less dense, permeate of all space, an intrinsic property of space and does not get diluted with the expansion of space. This fluid model of dark energy theory appears superior to the theory of modified gravity because of its strong evidence and calculative influence on the dynamics of the universe. But we can not deny the theory of modified gravity completely as we are ignorant about the secret behind gravity. Although the mathematical implications can be delicately designed that can match almost with the observed phenomenon of the universe but it may not be a real physical situation that can able to explain the different scenarios associated with this grand structure. Further, recent observational data \cite{8} supports dark energy model by predicting that nearly  68.3 \% of present universe is filled with this dark energy.

The unknown nature of dark energy evokes many of its forms like vacuum energy, phantom energy \cite{9,10} and quintessence \cite{11,12,13} etc which are characterized by their equation of state $p={\gamma_\phi}\rho$. Though vacuum energy, explained by cosmological constant, is the simplest one, but it suffers from fine-tuning problem: mismatching in the magnitude of cosmological constant as predicted by field theory and present observation by 123 order \cite{5}. So different kinds of dynamical dark energy models are purposed . Among them quintessence is a dynamical evolving dark energy with negative pressure, which in  cosmology is a real form of energy that is different from any normal matter or radiation, or even dark matter. This quintessence type of dark energy is characterized by its equation of state $p={\gamma_\phi}\rho$ with $0\ge$ ${\gamma_\phi} \ge-1$. Some other works \cite{bde,jq} have raised the possibility that the equation of state parameter $\gamma$ may be less than $-1$, which is known as phantom energy in literature. A peculiar property of cosmological models with phantom energy is the possibility of a Big Rip \cite{rrc}: an infinite increase of the scale factor of the Universe in a finite time. In the Big Rip scenario, the cosmological phantom energy density tends to infinity and all bound objects in the Universe are finally torn apart up to the subnuclear scales. But in some recent papers \cite{snso, snsd}, it is discussed that the final state of phantom Universe is not the Big Rip. They have considered that much before a Big Rip the quantum effects start to play the dominant role such that the universe undergoes a second quantum gravity era. There exist also ‘modified matter’ dynamical dark energy models like Chaplygin gas \cite{cg1,cg2}. But different type of dark energy models  introduce too many free parameters to able to fit with observational data or not able to explain all features of the universe, like for example coincidence problem: why the observed values of the cold dark matter density and dark energy density are of the same order of magnitude today although they differently evolve during the expansion of the universe \cite{ccp}. As a new alternative, different interacting dark energy models are discussed in literature.

Since a larger part of the universe is filled with dark energy and dark matter and there is lack of concrete knowledge about their origin and nature, we may assume that perhaps the dark energy and dark matter are coupled to each other giving rise to a single dark fluid.  Though this consideration seems slightly phenomenological, till date it is not rejected by any observation. Rather this idea is supported by particle physics, since dark energy and dark matter are thought to be of some fields and particle physics says that any two fields can interact with each other. Again the standard cosmological laws holds good when the interaction limit disappeared. Additionally the dark sector coupling can able to explain the cosmic coincidence problem. The idea of coupling in the dark sectors was initiated by Wetterich \cite{wetterich} and subsequently discussed by Amendola \cite{amendola} and others. In the couple of years, a rigorous analysis has been performed in the field of interacting dark energy by several authors with many interesting possibilities, see for instance \cite{a1,a2,a3,a4,a5,a6,a7,a8,a9,a10,a11,a12,14,bn}.

In our present study, we assumed a model where dark energy interacts with matter during evolution of the universe such that one may grow at the expense of other. Using this model, we try to study the expansion of the universe from very distant past to far future and explain the present day accelerated expansion of the universe.

\section{Basic Framework}
For a spatially flat $(k=0)$ FRW universe filled with dust and dark energy, the Friedmann equations take the form
\begin{equation} \label{fr1}
3H^2=8\pi G(\rho_\phi+\rho_m),
\end{equation}
\begin{equation} \label{fr2}
2{\frac{\ddot{a}}{a}}+H^2=-8\pi G p_\phi
\end{equation}
and the energy conservation equation becomes 
\begin{equation} \label{energy}
\dot\rho_m+\dot\rho_\phi+3H(\rho_m+\rho_\phi +p_\phi)=0.
\end{equation}
Where $H=\frac{\dot{a}}{a}$ is the Hubble parameter with $a$ as the scale factor, $\rho_\phi$ and $\rho_m$  are considered as the densities of dark energy and matter respectively and $p_{\phi}=\gamma_{\phi}\rho_{\phi}$ is the pressure of the dark energy with $\gamma_{\phi}$ as the equation of state parameter of dark energy. Assuming interaction between dark energy and matter, the energy conservation equation can be written as 
\begin{eqnarray} \label{int-rho}
\dot\rho_m+3H\rho_m=\delta
\end{eqnarray}
and 
\begin{eqnarray} \label{int-energy}
\dot\rho_\phi+3H\rho_\phi(1+\gamma_\phi)=-\delta=-\Gamma\rho_\phi
\end{eqnarray}
where $\delta=\Gamma\rho_\phi$ with $\Gamma$ as the interaction rate. 

In our model, we have assumed that the universe was started with only dust and dark energy appeared due to its decay with $\Gamma$ as the interaction rate. By considering observational data \cite{8} that at present $68.3$ percentage of the universe is filled with dark energy and the rest are matter which has been achieved in $13.82\times 10^9$ years through their interaction, we estimated that
$\Gamma \approx 1.566 \times 10^{-18} s^{-1}$. 

Again according to our model, the universe is filled with matter and dark energy. But standard model of cosmology, where dark energy is absent, assumes that the universe was radiation dominated upto a time $t_e$ where scale factor evolves like $a(t) \propto t^{\frac{1}{2}}$ and afterthat it is matter dominated where scale factor evolves like $a(t) \propto t^{\frac{2}{3}}$. Combining these two, here, we discuss the evolution of dark energy for two different era separately.

Integrating equation (\ref{int-energy}), one can easily get
\begin{eqnarray} \label{int1}
(ln {\rho_{\phi}})_{t<t_e}=-\frac{3}{2}(1+\gamma_\phi)ln t-\Gamma t +C_1
\end{eqnarray}
and
\begin{eqnarray} \label{int2}
(ln {\rho_{\phi}})_{t>t_e}=-2(1+\gamma_\phi)ln t-\Gamma t +C_2
\end{eqnarray}
where $C_1$ and $C_2$ are constants of integration and $t_e$ is the time of matter-radiation equality which is generally  taken as $t_e \sim 10^{11} s$.
\\ Simplifying above equations (\ref{int1}) and (\ref{int2}), we get
\begin{eqnarray} \label{rhor}
(\rho_\phi)_{t<t_e}=\rho_\phi(t_e)\bigg{(}\frac{t}{t_e}\bigg{)}^{-\frac{3}{2}(1+\gamma_\phi)}e^{-\Gamma(t-t_e)}
\end{eqnarray}
and
\begin{eqnarray} \label{rhom}
(\rho_\phi)_{t>t_e}=\rho_\phi(t_0)\bigg{(}\frac{t}{t_0}\bigg{)}^{-2(1+\gamma_\phi)}e^{-\Gamma(t-t_0)}
\end{eqnarray}
where $t_0$ is the present age of the universe having value $t_0 \sim 4.36126 \times 10^{17} s$.

\section{Deceleration Parameter and Present Universe}
The expansion of the universe is explained by Hubble's law. But whether the expansion is accelerating or decelerating one, is determined by deceleration parameter.
The deceleration parameter is defined as,
\begin{equation} 
q=-\frac{a\ddot{a}}{{\dot{a}}^2}
\end{equation}
Now equation (\ref{fr2}) can be written as
\begin{equation} \label{q}
q=\frac{1}{2}\bigg{(}1+\frac{8\pi G p_\phi}{H^2}\bigg{)}.
\end{equation}

Using equations (\ref{rhor}) and (\ref{rhom}) in equation (\ref{q}), we calculated the deceleration parameter for radiation-dominated and matter-dominated era as
\begin{equation} \label{qr}
q_{t<t_e}=\frac{1}{2}\bigg{[}1+32\pi Gt^2\gamma_\phi\rho_\phi(t_e)\bigg{(}\frac{t}{t_e}\bigg{)}^{-\frac{3}{2}(1+\gamma_\phi)}e^{-\Gamma(t-t_e)}\bigg{]}
\end{equation}
and
\begin{equation} \label{qm}
q_{t>t_e}=\frac{1}{2}\bigg{[}1+18\pi Gt^2\gamma_\phi\rho_\phi(t_0)\bigg{(}\frac{t}{t_0}\bigg{)}^{-2(1+\gamma_\phi)}e^{-\Gamma(t-t_0)}\bigg{]}.
\end{equation}

\begin{figure}[h]
\centering
\includegraphics[scale=0.4]{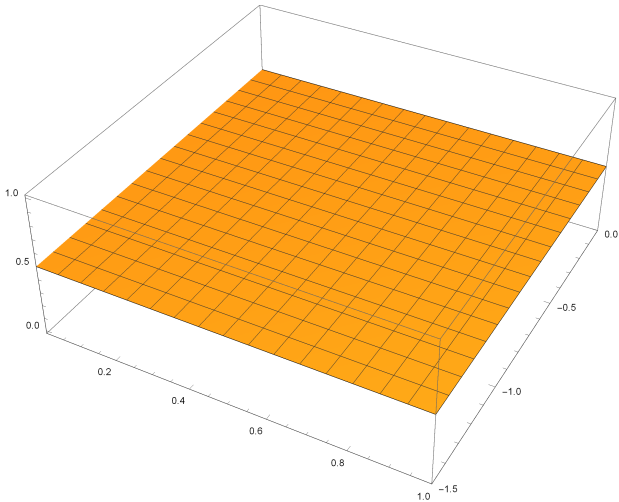}
\caption{Variation of deceleration parameter for different $\gamma_{\phi}$ varying from $-1.5$ to $0$ in radiation dominated era where $\frac{t}{t_e}$ varies between $0$ and $1$.}
\end{figure}

From equation (\ref{qr}), it can be easily found that in radiation-dominated era deceleration parameter $q$ takes only positive values i.e. $\frac{1}{2}$ as shown in Figure 1. On the other hand, equation (\ref{qm}) tells that in matter-dominated era, $q$ can take both positive and negative values. 
\\For present time $t=t_0$, above equation (\ref{qm}) gives

\begin{equation} \label{gamma}
\gamma_{\phi}=\frac{2q-1}{18 G \pi {t_0}^2 \rho_{\phi}(t_0)}
\end{equation}

Basing on the above equation (\ref{gamma}), we construct the Table I which gives the present values of equation of state parameters of dark energy for different values of deceleration parameter.
\begin{table}[ht]
\center
\caption{The present values of equation of state parameter of dark energy  $(\gamma_{\phi})$ for different values of deceleration parameter $(q)$ .}
\begin{tabular}{|c| c ||c| c |}\hline
 $q$ & $\gamma_{\phi}$ & $q$ & $\gamma_{\phi}$\\\hline 
-0.05 & -0.2042 & -0.55 & -0.3898\\
-0.10 & -0.2227 & -0.60 & -0.4083\\
-0.15 & -0.2413 & -0.65 & -0.4269\\
-0.20 & -0.2599 & -0.70 & -0.4455\\
-0.25 & -0.2784 & -0.75 & -0.4640\\
-0.30 & -0.2970 & -0.80 & -0.4826\\
-0.35 & -0.3155 & -0.85 & -0.5011\\
-0.40 & -0.3341 & -0.90 & -0.5197\\
-0.45 & -0.3527 & -0.95 & -0.5382\\
-0.50 & -0.3712 & -1.00 & -0.5568\\\hline
\end{tabular}
\end{table}

Comparing with the present observational data \cite{29} that the deceleration parameter is $-0.55$, the value of $\gamma_{\phi}$ for the present universe  is found from Table I to be $-0.3898$. In the early period of matter domination during which transition from deceleration to acceleration would occur, thus, the value of $\gamma_{\phi}$ is grater than (less negative) or equal to $-0.3898$. To showing this transition times for different equation of state parameters of dark energy $\gamma_{\phi}$, we construct the Table II and plot Figure 2 from equation (\ref{qm}). 
\begin{table}[ht]
\center 
\caption{Variation of transition time $(t_{q=0})$ and transition redshift $(z_{q=0})$ with equation of state parameters of dark energy  $(\gamma_{\phi})$.}
\begin{tabular}{|c |c |c |}\hline
$\gamma_{\phi}$ & Transition Time $(t_{q=0})$ & Transition Redshift $(z_{q=0})$\\\hline 
-0.3898 & 0.190$t_0 $ & 2.026\\
-0.35 & 0.182$t_0 $ & 2.114\\
-0.30 & 0.176$t_0 $ & 2.184\\
-0.25 & 0.180$t_0 $ & 2.137\\
-0.20 & 0.218$t_0 $ & 1.761\\
-0.18 & 0.276$t_0 $ & 1.359\\
-0.1661 & 0.472$t_0 $ & 0.649\\
-0.1660 & no transition & ----\\\hline
\end{tabular}
\end{table}

\begin{figure}[h]
\centering
\includegraphics[scale=0.6]{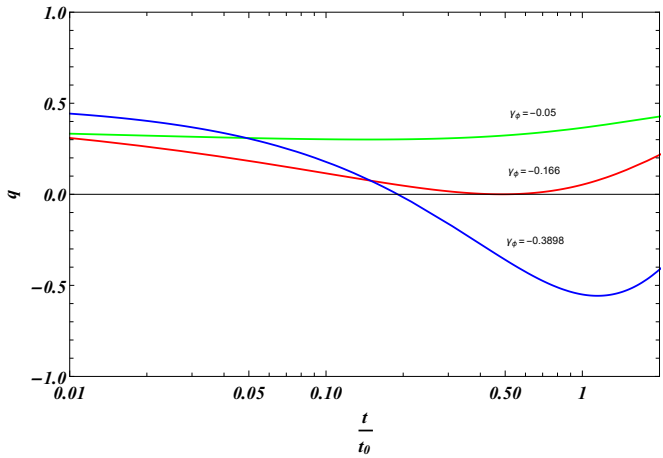}
\caption{Variation of deceleration parameter with time for different $\gamma_\phi$ in early matter-dominated era}
\end{figure}

From Table II and Figure 2, it is concluded that accelerated expansion is only possible for $\gamma_{\phi}$ less (more negative) than $-0.166$ and transition time would be in the range of $0.190 t_0$ to $0.472 t_0$. This gives a constraint on the transition redshift as $0.649 \le z_{q=0} \le 2.026$ which is in agreement with the observation \cite{30}.

\section{Nature of Future Expansion of the Universe}
Since at present time $\gamma_{\phi}=-0.3898$, for future expansion $\gamma_{\phi}$ must be less than (more negative) or equal to $-0.3898$. Considering the dark energy as quintessence type through out its evolution, we plot Figure 3 from equation (\ref{qm}) for showing variation of deceleration parameter with time for different $\gamma_{\phi}$.

\begin{figure}[h]
\centering
\includegraphics[scale=0.6]{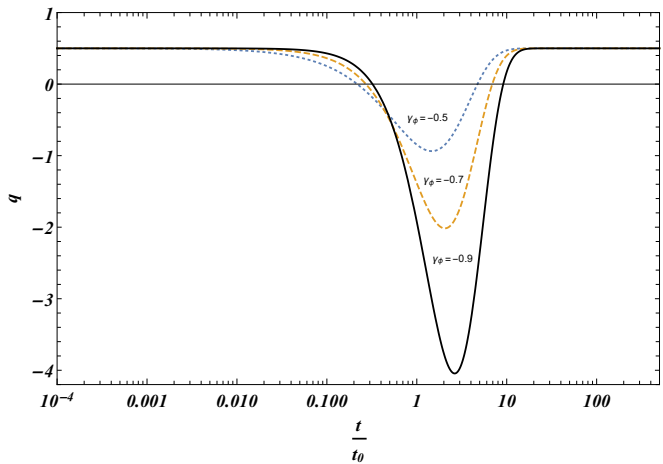}
\caption{Variation of deceleration parameter with time for different $\gamma_\phi$ of quintessence type dark energy}
\end{figure}

Figure 3 indicates that in near future the expansion of the universe will be undergo a second transition from accelerated to decelerated one and this deceleration will continues for far future. For getting this picture, we have assumed that the dark energy is quintessence type. But the same picture is repeated if we will assume the dark energy is phantom type i.e. $\gamma_{\phi} < -1$. In this regard, we construct Table III and plot Figure 4.\\

\begin{figure}[h]
\centering
\includegraphics[scale=0.6]{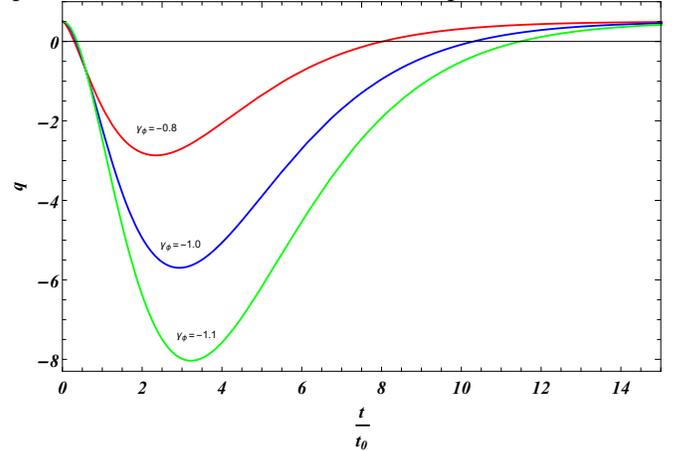}
\caption{Variation of deceleration parameter with time for different $\gamma_\phi$ in near future}
\end{figure}

\begin{table}[ht]
\center
\caption{Variation of second transition time $(t_{q=0})$ and saturation time $(t_{q=0.5})$ with different $\gamma_{\phi}$.}
\begin{tabular}{|c| c| c|}\hline
$\gamma_{\phi}$ & $t_{q=0}$ & $t_{q=0.5}$\\\hline 
-0.4 & 1.587 $\times 10^{18} s$ & 1.76 $\times 10^{19} s$\\
-0.5 & 2.061 $\times 10^{18} s$ & 1.82 $\times 10^{19} s$\\
-0.6 & 2.534 $\times 10^{18} s$ & 1.88 $\times 10^{19} s$\\
-0.7 & 3.020 $\times 10^{18} s$ & 1.94 $\times 10^{19} s$\\
-0.8 & 3.496 $\times 10^{18} s$ & 2.00 $\times 10^{19} s$\\
-0.9 & 3.989 $\times 10^{18} s$ & 2.06 $\times 10^{19} s$\\
-1.0 & 4.489 $\times 10^{18} s$ & 2.12 $\times 10^{19} s$\\
-1.1 & 5.000 $\times 10^{18} s$ & 2.18 $\times 10^{19} s$\\
-1.2 & 5.518 $\times 10^{18} s$ & 2.24 $\times 10^{19} s$\\
-1.3 & 6.044 $\times 10^{18} s$ & 2.30 $\times 10^{19} s$\\
-1.4 & 6.579 $\times 10^{18} s$ & 2.36 $\times 10^{19} s$\\
-1.5 & 7.121 $\times 10^{18} s$ & 2.42 $\times 10^{19} s$\\\hline
\end{tabular}
\end{table}

From Table III, we found that for equation of state parameter $\gamma_{\phi}$ becoming more and more negative, second transition time increases but in all casas deceleration parameter approaches a saturated value $\frac{1}{2}$ indicating a constant decelerated expansion in far future.
 
\section{Conclusion}
We, here, used an interacting dark energy model for explaining presently observed accelerated expansion of the universe. First we calculated the dark energy density parameter and then expressed deceleration parameter in terms of it for different epochs. In the next step, we have shown the variation of deceleration parameter $q$ with equation of state parameter $\gamma_\phi$. From our study, we found that in radiation-dominated era expansion of the universe is decelerating one since deceleration parameter $q$ is found to be always positive. But in matter-dominated era, the universe is changing gradually from a phase of decelerated to accelerated expansion as the calculated $q$ value shows a change from positive to negative. Comparing with the present observational data \cite{29} that the deceleration parameter is $-0.55$, the value of $\gamma_{\phi}$ for the present universe  is found to be $-0.3898$. The transition time at which the decelerated expanding universe switched to accelerated one, is found to be in the range of $0.189 t_0$ to $0.472 t_0$ which is in agreement with observation \cite{30}. Again we found that in near future the expansion of the universe will undergo a second transition from accelerated to decelerated one and this deceleration will continues for far future.

\noindent{\bf{Acknowledgements:}}
One of our authors Dr. Bibekananda Nayak is thankful to University Grants Commission, New Delhi, for providing financial support through UGC Start-Up-Grant Project having Letter No. F. 30-390/2017 (BSR).

\end{document}